\documentclass[final]{aipproc}

\layoutstyle{8x11single}

\begin{document}

\title[Enhancing non-Gaussianities]{Enhancing non-Gaussianities by\\ breaking local Lorentz invariance}


\author{Hael Collins}{
  address={The Niels Bohr International Academy, The Niels Bohr Institute, 2100 Copenhagen \O, Denmark}}

\begin{abstract}
This talk briefly explains how the breaking of a Lorentz-invariant description of nature at tiny space-time intervals might affect the non-Gaussian character of the primordial fluctuations left by inflation.  For example, a model that contains irrelevant operators that only preserve the spatial symmetries along constant-time surfaces can generate a larger non-Gaussian component in the pattern of primordial fluctuations than is ordinarily predicted by inflation.  This property can be useful for constraining models that allow some Lorentz violation at short distances, beyond the constraints possible from the power spectrum alone.
\end{abstract}

\maketitle

\section{Introduction}

The universe today is filled with a quite bewildering variety of structures, from stars and galaxies at comparatively small scales (when judged by the size of the observable universe) to the vast clusters and networks of galaxies and emptier voids among them at the largest scales.  But the earlier universe differs markedly from its current appearance.  The earliest relics that we can see directly show a remarkable uniformity in how the material was distributed at those times.  For example, the temperature of the radiation that first escaped as the universe cooled from a hot plasma into a transparent gas varies less than one part in one hundred thousand from one place to another.  Moreover, the abundances of the primordially produced lighter elements also seem to be the same everywhere in the universe, suggesting that this uniformity extends to still earlier times.  Yet the universe needed to have had at least {\it some\/} variation, even at the very earliest of times, for without any tiny spatial variations, the growth of the structures that we see today could never have begun in the first place.

The theory of inflation provides one elegant mechanism for explaining the origin of these tiny {\it primordial fluctuations\/}.  One very appealing property of the inflationary picture is that such fluctuations are unavoidable, being a simple consequence of the two basic ingredients needed in any inflationary model.  A typical model for inflation requires a space-time that is expanding at an accelerating rate and a quantum field whose dynamics are responsible for the expansion.  But a quantum field is always fluctuating.  It would be quite impossible for it to be otherwise, as it would be inconsistent with the Heisenberg principle for a field to be at once stationary and without any spatial fluctuations.  Let us describe this behavior a little more precisely.  If we denote the quantum field by $\varphi(t,\vec x)$ and its quantum state by $|0(t)\rangle$, then while its mean value in this state may vanish on average, 
\begin{equation}
\langle 0(t)| \varphi(t,\vec x) |0(t)\rangle = 0 ,
\label{onepoint}
\end{equation}
the variance of the field never does, 
\begin{equation}
\langle 0(t)| \varphi(t,\vec x) \varphi(t,\vec y) |0(t)\rangle \not= 0 . 
\label{variance}
\end{equation}
Of course, such quantum fluctuations are also occurring at tiny scales today.  The difference between the early universe and the universe today is that, during the former, the space-time expands very rapidly, stretching these tiny fluctuations to very large scales---even as large as the observable universe today.

Because these fluctuations are such a basic consequence of inflation, the pattern that inflation predicts for them is fairly insensitive to the details of a particular model, though measurements have actually reached a precision that can test inflation beyond its most basic predictions and can even exclude individual models.  Nevertheless, the overall inflationary picture is still in a superb agreement with what is inferred from observations.  Inflation predicts 
\begin{itemize}
\item that the fluctuations have more or less the same amplitude at all length scales---including those that otherwise would not have been in causal contact without an inflationary period, 
\item that the fluctuations are all in phase and are adiabatic, 
\item that their pattern is largely a {\it Gaussian\/} one, the meaning of which we shall explain later, and
\item that there should also be a background of primordial gravity waves.
\end{itemize}
Each of these predictions---except for the last---matches what is seen in all experiments so far.

With this success, it is important to test the underlying assumptions of the inflationary picture further and to shed some light on its otherwise mysterious elements.  What exactly is this field that is producing the inflation and how does it fit into the rest of our picture for particle physics?  How did the universe find itself initially in a spatially flat configuration, at least to a sufficient degree over a sufficiently large patch?  Does the size of the field change rapidly over small enough distances that we ought to worry about quantum gravity?  Which is the correct quantum state $|0(t)\rangle$ to choose?  And is it possible to have {\it too much\/} expansion of the quantum fluctuations for the consistency of our picture?

If these questions seem to be only so much cavilling, it instructive to compare the setting used in inflation with that of ordinary particle physics to realize how much less we know about the former and how much more limited is our ability to test it.  In particle physics,
\begin{itemize}
\item We know the background (flat space).
\item We know all the relevant ingredients up to some energy scale (and any unseen stuff remains largely decoupled from the lower energy world).
\item We have a well developed, {\it and well tested\/}, theoretical framework (quantum field theory in flat space).
\item Gravity is entirely negligible.
\item And perhaps most importantly, we are free to test anything we want, in a controlled environment, up to a limiting energy scale.
\end{itemize}
Now contrast each of these points with what we assume for the early universe. 
\begin{itemize}
\item We probably know the background (it seems to be approaching a spatially independent one at early enough times).
\item We can only guess what are relevant ingredients, and have no way to observe them directly (at least so far, and probably for a long while yet to come).
\item We have a framework, quantum field theory in curved space; and while it is generally self-consistent, it has not been tested.
\item Gravity is an essential ingredient.
\item And experimentally, we can only observe a very limited and indirect set of evidence, and we are not able to test the picture directly (at least not to the standard applied in an accelerator experiment).
\end{itemize}
Nothing on this list implies that the inflationary picture is necessarily wrong; but each item does indicate some unresolved piece of the inflationary framework which we would like to understand better.  Because of its remoteness from anything that we can test directly and because it will probably be difficult to find {\it complementary\/} observations to those that we already have which can test this picture, it is important to address some of these questions if we are to have a little more confidence that inflation really did occur in our own universe.

\section{The primordial fluctuations}

Let us return to the primordial fluctuations, this time in more depth.  In the inflationary picture, they correspond to the fluctuations in the space-time background itself---that is, variations in the strength of gravity from one place to another.  Since every other ingredient of the universe feels the force of gravity, these space-time fluctuations are transferred into fluctuations in the distribution of matter and radiation, which in turn affect something that we can observe.  Such fluctuations, for example, produce a very characteristic pattern in the temperature and the polarization of the light of the cosmic background radiation mentioned earlier, or in how galaxies are distributed over very large scales.  By measuring these patterns very precisely, we try to infer something about the quantum state of the universe at a much earlier epoch.

A fluctuation generated by inflation, which we shall call $\zeta(t,\vec x)$, is a linear combination of the scalar field responsible for the inflation and a part of the fluctuations of the space-time metric.  The classical background is assumed to be spatially invariant
\begin{equation}
ds^2 = dt^2 - a^2(t)\, d\vec x\cdot d\vec x . 
\label{bkgrmetric}
\end{equation}
Since any cosmological observation measures a classical quantity, we never see $\zeta(t,\vec x)$ directly as a quantum field.  Instead, we first convert it into a set of classical functions by taking the expectation value of a product of the fields, 
\begin{equation}
C_n(t; \vec x_1, \ldots, \vec x_n) = 
\langle 0(t)| \zeta(t,\vec x_1) \cdots \zeta(t,\vec x_n) |0(t)\rangle . 
\label{npoint}
\end{equation}
The $|0(t)\rangle$ is the quantum state of the field; choosing it almost always requires making some assumptions about the properties and symmetries of the universe deep within the inflationary era.

From the perspective of the post-inflationary stages of cosmology, the functions $C_n(t; \vec x_1, \ldots, \vec x_n)$ provide the ``initial conditions'' needed to solve completely how the all of the components of the universe evolve.  The dynamics of these ingredients---gravity, dark matter, baryonic matter, radiation, neutrinos, etc.---are governed by a coupled system of equations derived from Einstein's equations and a Boltzmann equation for each of the species.  In the beginning, the perturbations for each ingredient about its own smooth background are quite small, so during this early stage the coupled equations can be treated to linear order to yield relatively simple predictions for the appearance of the universe at later stages, once we have specified some particular choice for the ``initial conditions'' for these equations

Because the fluctuations of the background radiation or in the distribution of distant galaxies are so small---and observing them is so difficult---only the lowest order correlation functions for the temperature of the radiation or the distribution of matter can be measured with a sufficient precision.  In turn, such relatively crude measurements allow only the very simplest properties of the primordial fluctuations to be inferred.  So far only the two-point function, 
\begin{equation}
\langle 0(t)| \zeta(t,\vec x) \zeta(t,\vec y) |0(t)\rangle , 
\label{twopoint}
\end{equation}
has been unambiguously measured.  Both its overall size as well as---to a much lesser extent---its basic dependence on the scale $|\vec x-\vec y|$ have been found by fitting cosmological models for the evolution of the early universe against observations.  Even the next largest, the three-point function, 
\begin{equation}
\langle 0(t)| \zeta(t,\vec x)\zeta(t,\vec y)\zeta(t,\vec z) |0(t)\rangle , 
\label{threepoint}
\end{equation}
has not been seen yet, though the bounds on its possible size are ever improving.  Measuring---or even just sufficiently constraining---the size of this function would provide an extremely useful test of the inflationary picture, since most inflationary models predict a pattern of fluctuations that is highly {\it Gaussian\/} in its character.

A {\it purely\/} Gaussian pattern is one where all of the information is captured by just the two-point function.  All higher order $n$-point functions either vanish, if $n$ is odd, or can be expressed as a sum of products of two-point functions, if $n$ is even.  For example, for a purely Gaussian pattern, the four-point function would be  
\begin{eqnarray}
\langle 0| \zeta(t,\vec x_1) \zeta(t,\vec x_2) \zeta(t,\vec x_3) \zeta(t,\vec x_4) |0\rangle 
&\!\!\!\!=\!\!\!\!& 
\langle 0| \zeta(t,\vec x_1) \zeta(t,\vec x_2)|0\rangle\,  
\langle 0| \zeta(t,\vec x_3) \zeta(t,\vec x_4) |0\rangle 
\nonumber \\
&& 
+\ \langle 0| \zeta(t,\vec x_1) \zeta(t,\vec x_3)|0\rangle\,  
   \langle 0| \zeta(t,\vec x_2) \zeta(t,\vec x_4) |0\rangle 
\nonumber \\
&& 
+\ \langle 0| \zeta(t,\vec x_1) \zeta(t,\vec x_4)|0\rangle\,  
   \langle 0| \zeta(t,\vec x_2) \zeta(t,\vec x_3) |0\rangle . 
\label{fourpoint}
\end{eqnarray}
However, inflation is never expected to produce such an ideally Gaussian pattern, but only a pattern that is nearly so.  Signals of any non-Gaussianity, such as the three-point function should then be small to a degree which we shall define.

The action for the quantum fluctuations is derived by starting from the Einstein action, plus a simple action for the scalar field $\Phi(t,\vec x)$ whose dynamics are responsible for the inflationary expansion, 
\begin{equation}
\int d^4x\, \sqrt{-g} \Bigl\{ {\textstyle{1\over 2}} M_{\rm pl}^2 R 
+ {\textstyle{1\over 2}} g^{\mu\nu} \partial_\mu\Phi\partial_\nu\Phi
- V(\Phi) \Bigr\} . 
\label{fullaction}
\end{equation}
Here, $R$ is the Riemann scalar curvature for the metric $g_{\mu\nu}$ and $V(\Phi)$ is the potential for the scalar field.  The scale $M_{\rm pl}$, called the {\it Planck mass\/}, is an important one, for it corresponds to the mass scale at which quantum gravity can no longer be treated perturbatively, even as an effective theory.  Therefore, it is important for the consistency of our treatment that we restrict to situtations that do not depend sensitively on what happens near or beyond this scale.

Inflation assumes that the universe can be described by a classical, spatially constant background,
\begin{eqnarray}
ds^2 &\!\!\!\!=\!\!\!\!& dt^2 - a^2(t)\, d\vec x\cdot d\vec x ,
\nonumber \\
\phi(t) &\!\!\!\!=\!\!\!\!& \langle 0(t) | \Phi(t,\vec x) |0(t)\rangle , 
\label{background}
\end{eqnarray}
about which there are small quantum fluctuations, $\zeta(t,\vec x)$.\footnote{More specifically \cite{perts,bardeen}, if we add quantum fluctuations to both the scalar field, $\Phi(t,\vec x) = \phi(t) + \delta\phi(t,\vec x)$ and the spatial components of the metric, $g_{ij}(t,\vec x) = a^2(t) [-1+2\Psi(t,\vec x)]\, \delta_{ij}$, then $\zeta = \Psi + (H/\dot\phi)\delta\phi$.}  A natural dynamical scale is provided by the rate at which the space-time is expanding, 
\begin{equation}
H(t) = {1\over a(t)} {da\over dt} = {\dot a\over a} , 
\label{hubbledef}
\end{equation}
and it is often called the {\it Hubble scale\/}.  Starting from the action above and laboriously expanding in powers of $\zeta(t,\vec x)$, we eventually find that the action for the fluctuations \cite{perts},
\begin{equation} 
S[\zeta] = S_0[\zeta] + S_I[\zeta] , 
\label{zaction}
\end{equation}
can be divided into a quadratic piece,  
\begin{equation}
S_0 = \int dt\, a^3\, {\dot\phi^2\over H^2} 
\int d^3\vec x \biggl\{ {1\over 2} \dot\zeta^2 
- {1\over 2} {1\over a^2} \vec\nabla\zeta\cdot \vec\nabla\zeta \biggr\} , 
\label{S0}
\end{equation}
and an interacting piece $S_I$ that contains all of the cubic and higher order interactions.

Varying the quadratic action with respect to the field $\zeta$ yields its free equation of motion, 
\begin{equation}
{d\over dt} \biggl\{ a^3\, {\dot\phi^2\over H^2} \dot\zeta \biggr\} 
- a\, {\dot\phi^2\over H^2} \vec\nabla\cdot\vec\nabla\zeta = 0 , 
\label{eom}
\end{equation}
which we can solve completely once we treat $\zeta(t,\vec x)$ as a canonically normalized quantum field and have added one further boundary condition.  The usual choice for this boundary condition is expressed as a condition on the short-distance behavior of the fields---or, if we expand the field in its Fourier modes, 
\begin{equation}
\zeta(t,\vec x) = \int {d^3\vec k\over (2\pi)^3}\, 
\bigl\{ \zeta_k(t) e^{i\vec k\cdot\vec x} a_{\vec k} 
+ \zeta_k^*(t) e^{-i\vec k\cdot\vec x} a_{\vec k}^\dagger \bigr\} , 
\label{fourier}
\end{equation}
as a condition on its high-spatial momentum behavior.  The $a_{\vec k}^\dagger$ and $a_{\vec k}$ here are creation and annihilation operators.  If we define an initial state for the fluctuations, $|0\rangle = |0(t_0)\rangle$, at some initial time $t_0$, then $a_{\vec k}\, |0\rangle=0$.  So the choice of the mode functions, $\zeta_k(t)$, since they are associated with a particular way of dividing the creation operators from the annihilation operators, equally corresponds to a choice of the initial state of the field.

The standard choice is to use the state that most closely resembles the vacuum state of flat space.  Since at large distances or time-intervals, an inflationary universe looks nothing like flat space, this condition properly applies to only very short distances, which are those for which $k\gg a(t_0)H(t_0)$.  This condition then selects the state with the following associated Fourier modes, 
\begin{equation}
\zeta_k(t) = {H\over\dot\phi} {H\over M_{\rm pl}} {1\over\sqrt{2k^3}} 
\biggl( i + {k\over H} e^{-Ht} \biggr) \exp\biggl[ i {k\over H} e^{-Ht} \biggr] 
+ \cdots , 
\label{zetak}
\end{equation}
to leading order in an expansion that we define next.

Solving for the full time dependence of the theory is usually not possible for a general inflationary background, $\big\{ a(t), \phi(t) \bigr\}$.  However, to have had enough inflation that all of the observable universe today was once in causal contact already places some fairly strong constraints on the possible behavior of the scalar field, namely, that its derivatives cannot be too large.  These constraints are often called the {\it slow-roll conditions\/}.  So in writing the modes thus, we have implicitly neglected any corrections that are small in this slowly rolling limit.  These conditions can be expressed by introducing two dimensionless parameters for the first and second derivatives of $\phi$, 
\begin{equation} 
\epsilon \equiv {1\over 2} {\dot\phi^2\over H^2}
\qquad\hbox{and}\qquad
\delta \equiv {\ddot\phi\over H\dot\phi} , 
\label{slowroll}
\end{equation}
which should then both be small.  One of these parameters appears in the mode functions, which can be rewritten as 
\begin{equation}
\zeta_k(t) = {1\over 2} {1\over\sqrt{\epsilon}} 
{H\over M_{\rm pl}} {1\over k^{3/2}} 
\biggl( i + {k\over H} e^{-Ht} \biggr) \exp\biggl[ i {k\over H} e^{-Ht} \biggr] 
+ \cdots . 
\label{zetakeps}
\end{equation}
In the slowly-rolling limit, the background expands essentially at an exponential rate, 
\begin{equation}
a(t) = e^{Ht} + \cdots . 
\label{scale}
\end{equation}

We have finally enough information to solve for the inflationary prediction for the leading behavior of the two-point function, 
\begin{equation}
\langle 0(t)| \zeta(t,\vec x) \zeta(t,\vec y) |0(t)\rangle 
= \int {d^3\vec k\over (2\pi)^3}\, e^{i\vec k\cdot(\vec x-\vec y)} 
\zeta_k(t)\zeta_k^*(t) + \cdots .
\label{twopointlead}
\end{equation}
The modes that are relevant for the observations today are those that have been stretched to a large size (or small wavenumber $k$) by the end of inflation, which means in practice that
\begin{equation}
k \ll a(t)H(t)
\qquad\Rightarrow\qquad 
e^{-Ht}k \ll H . 
\label{outofH}
\end{equation}
Applying this limit yields the two-point function for the pattern of primordial fluctuations produced by inflation, 
\begin{equation}
\langle 0(t)| \zeta(t,\vec x) \zeta(t,\vec y) |0(t)\rangle 
= {1\over 8\pi^2} {1\over\epsilon} {H^2\over M_{\rm pl}^2} 
\int_0^\infty {dk\over k}\, 
{\sin\bigl( k|\vec x-\vec y| \bigr)\over k |\vec x-\vec y| }
+ \cdots .
\label{2point}
\end{equation}
We can now understand better what is meant by a ``small'' set of primordial fluctuations:  the dimensionless prefactor should itself be small.  This condition implies that the inflationary scale $H$ ought to be well below the Planck scale $M_{\rm pl}$, which is good for the consistency of the picture.  Notice, moreover, that the integral has no preferred scale; were we to simultaneously rescale $k\to \lambda k$ and $|\vec x - \vec y| \to \lambda^{-1}|\vec x - \vec y|$, it would be unchanged.  Had we included the next set of corrections to the two-point function in the slowly rolling limit, we would have found that it does acquire a slight scale dependence.

Unlike the two-point function whose scale dependence is already beginning to be accessible to observations, even the correct size of the three-point function is still unknown empirically.  Sometimes a crude measure is used to parameterize its size, allowing different models to be compared on a similar footing.  For example, consider a model\footnote{Since the fluctuations mentioned in this paragraph are not exactly the same as those in the rest of this article, we shall write them with a tilde to keep them distinct from the rest.} where we can assume that the true field $\tilde\zeta(t,\vec x)$ can be approximately related to a purely Gaussian one $\zeta_g(t,\vec x)$ through a non-linear field redefinition of the form,
\begin{equation}
\tilde\zeta(t,\vec x) = \zeta_g(t,\vec x) 
- {\textstyle{3\over 5}} f_{\rm nl}\, \bigl[ \zeta_g^2(t,\vec x)
- \langle 0(t)| \zeta_g^2(t,\vec x) |0(t)\rangle \bigr] , 
\label{nonlinear}
\end{equation}
the factor of ${3\over 5}$ being a commonly used convention.  We can then evaluate the size of this three-point function (not the true three-point function) in terms of the parameter $f_{\rm nl}$.  The expectation value has been subtracted in the second term to keep the one-point function of $\tilde\zeta(t,\vec x)$ equal to zero.  Because of the quadratic piece, the three-point function of $\tilde\zeta(t,\vec x)$ can be expanded in terms of two-point functions of the Gaussian field
\begin{eqnarray}
\langle 0(t)| \tilde\zeta(t,\vec x) \tilde\zeta(t,\vec x) \tilde\zeta(t,\vec x) |0(t)\rangle 
&\!\!\!\!=\!\!\!\!& 
-\ {\textstyle{6\over 5}} f_{\rm nl}\, 
\langle 0| \zeta_g(t,\vec x)\zeta_g(t,\vec y) |0\rangle 
\langle 0| \zeta_g(t,\vec x)\zeta_g(t,\vec z) |0\rangle 
\nonumber \\
&&
-\ {\textstyle{6\over 5}} f_{\rm nl}\, 
\langle 0| \zeta_g(t,\vec x)\zeta_g(t,\vec y) |0\rangle 
\langle 0| \zeta_g(t,\vec y)\zeta_g(t,\vec z) |0\rangle 
\nonumber \\
&&
-\ {\textstyle{6\over 5}} f_{\rm nl}\, 
\langle 0| \zeta_g(t,\vec x)\zeta_g(t,\vec z) |0\rangle 
\langle 0| \zeta_g(t,\vec y)\zeta_g(t,\vec z) |0\rangle 
+ \cdots . 
\label{3pointzetatilde}
\end{eqnarray}
Now, the mode functions that we derived earlier were those for the quadratic or {\it Gaussian\/} part of the action, so we can immediately use them here, introducing some $\delta$-functions along the way which let us combine the two separate momentum integrals for each pair of two-point functions into a single integral over three spatial momenta, obtaining 
\begin{eqnarray}
&&\!\!\!\!\!\!\!\!\!\!\!\!\!\!\!\!\!\!\!\!\!\!\!\!
\langle 0(t)| \tilde\zeta(t,\vec x) \tilde\zeta(t,\vec x) \tilde\zeta(t,\vec x) |0(t)\rangle 
\nonumber \\
&&
= - {3\over 5} f_{\rm nl} {1\over\epsilon^2} {H^4\over M_{\rm pl}^4} \, 
\int {d^3\vec k_1\over (2\pi)^3} {d^3\vec k_2\over (2\pi)^3} 
     {d^3\vec k_3\over (2\pi)^3}\, 
e^{i\vec k_1\cdot\vec x} e^{i\vec k_2\cdot\vec y} e^{i\vec k_3\cdot\vec z} 
(2\pi)^3\, \delta^3(\vec k_1 + \vec k_2 + \vec k_3)
{k_1^3+k_2^3+k_3^3\over 8 k_1^3k_2^3k_3^3} 
+ \cdots . 
\label{fnldef}
\end{eqnarray}
Note that we have already taken the limit where the modes have all been stretched to a large size, $k_i \ll a(t)H(t)$.  The reason that we have been carefully distinguishing these toy fluctuations $\tilde\zeta(t,\vec x)$ from the genuine ones is that the momentum dependence of the true three-point function for $\zeta(t,\vec x)$ will be different, in general.  Nevertheless, we can still match match this toy three-point function with the true one in particular corners of momentum space---when $k_1\sim k_2\sim k_3$ are all nearly equal, for instance---to define an ``effective $f_{\rm nl}$'' for the true theory.

\section{The trans-Planckian problem}

Until now, we have been describing the inflationary picture without worrying too much about the questions that we mentioned earlier when we compared inflation with particle physics in flat space.  We shall next look a little more closely at some of the assumptions that we have been making to learn whether they imply any limitations to our picture.  Recall that for the treatment to remain consistent, we are limited to distances or time intervals that are larger than a Planck length, $M_{\rm pl}^{-1}$.  The reason for doing so is that the theory of gravity that we have been using receives large quantum corrections to processes when we evaluate them at these tiny scales.  For example, we evaluated the two-point function earlier by assuming that its largest contribution comes from the quadratic terms included in $S_0 $ and that everything else is smaller and can therefore be treated perturbatively.  Although we did not show the details of its derivation here, the action that we used to find $S_0$ (and $S_I$) should be seen as only the first part of a larger one; quantum corrections, for example, can generate terms of the form,
\begin{equation}
\int d^4x\, \sqrt{-g}\, \bigl\{ a\, R^2 + b\, R_{\mu\nu}R^{\mu\nu} 
+ c\, R_{\mu\nu\lambda\sigma}R^{\mu\nu\lambda\sigma} 
+ {\cal O}(R^3) \bigr\} ,
\label{Rsquared}
\end{equation}
even if they were not originally present, which will generally alter the form of the quadratic action $S_0$ for $\zeta(t,\vec x)$ at distances shorter than the Planck length, $M_{\rm pl}$.  If the metric changes rapidly enough over this scale, then the two derivatives in the scalar curvature will yield, $R\sim \partial\partial g \sim M_{\rm pl}^2$, and the terms here suddenly become of an equal importance with the Einstein-Hilbert term, ${1\over 2} M_{\rm pl}^2 R$, in our former action.

Moreover, our framework also assumed that we could more or less treat the quantum fluctuations as though they existed in an essentially fixed, classical background,
\begin{equation}
ds^2 = dt^2 - a^2(t)\, d\vec x\cdot d\vec x . 
\label{metricagain}
\end{equation}
But at distances smaller than a Planck length, the fluctuations (which were partially the quantum fluctuations of the space-time) could be large enough that this simple division of the space-time into a spatially independent, classical part and a spatially varying quantum part might no longer hold.  So for a consistent picture---at least without adding some new principle or ingredient---we ought to avoid calculations of observable quantities that depend on evaluating something in this Planckian regime.

Of course, none of these arguments are unique to an inflationary background; they apply equally to flat space.  But in flat space we have a natural notion of {\it decoupling\/}, the idea that phenomena at disparate scales should have relatively little influence on each other.  In a sense, quantum gravitational corrections for ``bare'' processes in flat space can be large.  But since these effects only arise from the behavior at extremely small length or time scales, they can be removed by rescaling the parameters in the action, adding the appropriate set of local operators, to define a ``physical'' version of the theory.  This physical theory is finite, calculable and predictive and it is insensitive to the details at the Planck scale.  In the process, we can no longer regard the parameters as fixed constants of nature; instead, they vary with the scale that we are investigating and they need to be defined by applying an appropriate renormalization condition.  Moreover, in flat space there is at least one globally defined frame in which anything with a wavelength larger than a Planck length remains so, as long as it is free of any other influences.  The important difference from this setting is that in an inflationary background the space-time itself expands---and at a very dramatic rate.  Setting a cut-off for short distances, for instance, becomes a condition that depends on time.

The trouble with an expanding background is that any fluctuation was inevitably smaller in the past.  To take a definite example, consider take the wavelength of a fluctuation responsible for some feature in the cosmic background radiation, $a(t_{\rm now})/k$.  We can then ask how big was its wavelength during some early stage of inflation, $a(t)/k$.  Inflation is primarily a way for stretching the radius of a causal patch of the universe until what we see today could all have once been within the same Hubble horizon $H^{-1}$.  So at a minimum, all the wavelengths ($k$'s) relevant to cosmological observations ought to satisfy 
\begin{equation}
{a(t)\over k} < {1\over H(t)} 
\qquad{\rm or}\qquad 
k > a(t) H(t) . 
\label{scales}
\end{equation}
Moreover, $t$ might not be the beginning of inflation; if we look still further back to $t' < t$, then the wavelength will only have been the smaller, by a factor of $a(t')/a(t)$.  With {\it enough\/} inflation, we inevitably find that modes being used to make predictions about cosmological measurements today were at one time the fluctuations of a quantum field over distances smaller than a Planck length.  This observation is called the {\it trans-Planckian problem of inflation\/} \cite{transPlanck,picture}.

Let us look at this problem from a more pictorial perspective.  Suppose that we follow a mode $\zeta_k(t)$ back in time.  Its physical wavelength at any time $t$ is given by $a(t)/k$.  To be more definite, and to choose something quite large on the scale of the universe today, we examine a mode that became smaller than the Hubble horizon $H^{-1}(t)$ at a time during the era when matter was the most abundant ingredient.  During the matter era---and the radiation era before it---the horizon grows faster than the rate at which the wavelength of a fluctuation grows.  In contrast, during the inflationary era this ordering is dramatically reversed---the wavelength grows exponentially while the horizon hardly changes at all.  The more detailed time dependence of the wavelength associated with $\zeta_k(t)$ and the Hubble horizon for each of these epochs is shown in the following table:
\vskip6truept

\begin{table}[h]
\begin{tabular}{c|cccc}
\hline 
& &inflationary &radiation &matter \\
& &era          &era       &era    \\
\hline
wavelength &$a(t)/k\quad\propto$ &$e^{Ht}$ &$t^{1/2}$ &$t^{2/3}$ \\
\hline
horizon    &$1/H(t)\quad\propto$ &constant &$2t$      &${3\over 2}t$ \\
\hline
\end{tabular}
\end{table}

\vskip6truept\noindent
Plotting the physical size of the wavelength, $a(t)/k$, of our chosen mode over time, we find the following picture.\footnote{This picture is similar to one appearing in \cite{picture}.  Note that the plot for the wavelength of the mode is only meant to be illustrative, not exact.} 
\begin{figure}[h]
\includegraphics{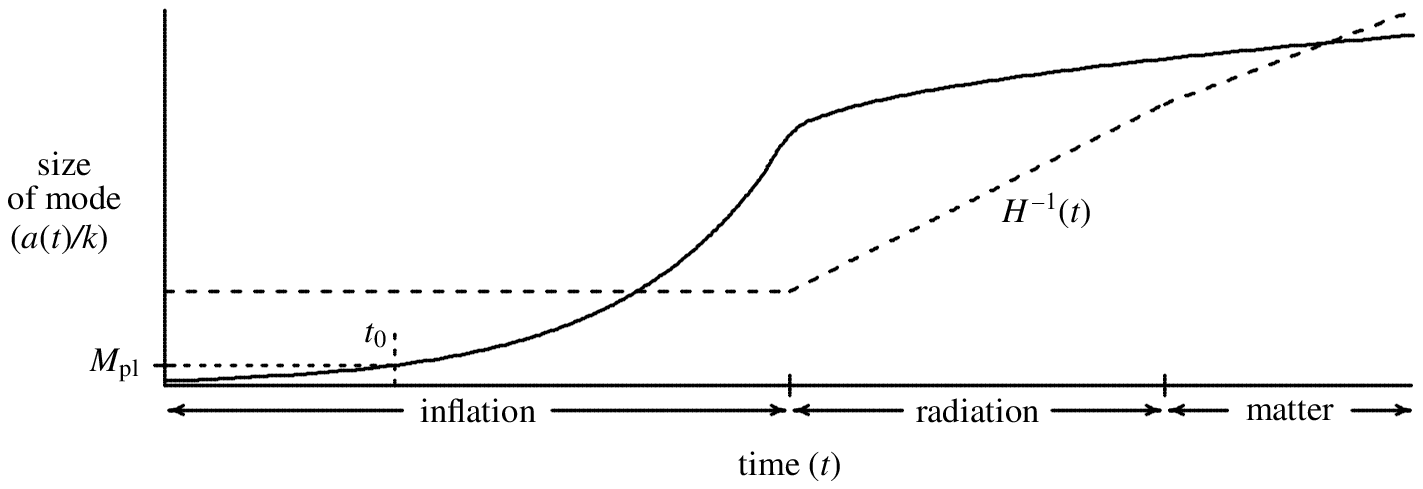}
\end{figure}

\noindent
The dashed line shows the evolution of the Hubble horizon $H^{-1}(t)$ for comparison.  At the far right end, the wavelength is within the horizon of the universe, so the fluctuation $\zeta_k(t)$ can influence how matter is distributed at that scale.

We now follow the mode backwards in time by reading the picture from right to left.  A little earlier, we see that the mode had not yet reentered the horizon.  For a long while, it was beyond the influence of any process, other than the expansion of the space-time.  For the particular mode that we have drawn, this stage lasted during the first part of the matter era, throughout the entirety of the radiation era, and during the latter part of the inflationary era.  Farther back, during inflation there was a moment when the mode was first leaving the horizon.  Before this instant, it was still possible for the mode to be subject to causal processes.  However, if we look even further back, by exactly another $\ln(M_{\rm pl}/H)$ ``$e$-folds,'' the wavelength of this mode would have been equal to the Planck length, at a time we have labeled as $t_0$ in the picture---and prior to $t_0$ the wavelength was still smaller.  Other modes, corresponding to different values of $k$, will be shifted up or down relative to the one we have drawn, but their histories are essentially the same.

It might be that nothing at all odd happens as we cross the Planck threshold and that the modes $\zeta_k(t)$ remain exactly as we chose them; however, this situation would imply strong constraints on the quantum behavior of gravity and that space-time is much tamer at these scales than is usually envisaged.  Alternatively, some principle or mechanism might cause {\it apparently\/} asymptotically flat space modes naturally to emerge once the physical wavelength of a mode has been stretched larger than a Planck length, though such a mechanism has not yet been discovered.

\section{Non-Gaussianities}

One method for constraining what could happen as we approach this Planckian threshold is to relax one of the properties that we usually assume about the universe at those scales and evaluate the consequences for observations.  In doing so, we must still avoid the trans-Planckian regime ourselves.  We therefore introduce an ``initial time'' $t_0$ into our treatment.  This time is not meant to be the beginning of inflation; rather, it is simply the earliest time that can still be used to calculate some particular observable without any of its relevant length scales ever having been smaller than a Planck length.  The prior history, or the properties of nature beyond the Planck scale, is encoded in the choice for the initial quantum state of the fluctuations, $|0\rangle = |0(t_0)\rangle$, or in the set of operators consistent with the new properties of the universe that we have allowed at trans-Planckian scales.  In essence, our treatment is an {\it effective\/} one, in the sense that though we forgo explaining the full history of the universe, we at least stay within a regime where we can trust our calculational framework.

To take a concrete example, let us return to the non-Gaussianities.  The standard prediction for inflation is that the dimensionless parameter $f_{\rm nl}$---introduced earlier as a measure of the amplitude of the three-point function---should be roughly of the same size as the slowly rolling parameters \cite{threepoint}, 
\begin{equation}
f_{\rm nl} \sim {\cal O}(\epsilon, \delta) \ll 1 . 
\label{stndfnl}
\end{equation}
We can study what would happen to this prediction if one of the properties of nature were very different from the ones that we have been assuming.  For instance, suppose that on scales smaller then a Planck length, the background is not described well by a space-time which is approximately flat.  To keep the analysis general, we do not make any specific assumptions about how nature behaves at these scales, but instead add all the allowed operators to the action that are consistent with the broken local Lorentz symmetry.  Here for illustration we break just the symmetries between the time and space directions, preserving the invariance under spatial translations and rotations.  This symmetry-breaking should be implemented so that it only affects the behavior at short-distances, for otherwise it would already have been observed. 

We can introduce this symmetry-breaking in one of two complementary ways \cite{iscap,effstate}.  We can either alter the structure of the quantum state of the fluctuations $|0\rangle$ by altering the form of the Fourier modes $\zeta_k(t)$ beyond some threshold $k$, or, to take a more conventional approach, we could add new {\it irrelevant\/} operators that reflect the diminished symmetry that we are assuming for those scales.  Irrelevant operators have a mass dimension greater than four---they have negligible effects at larger distances but they become important once we reach sufficiently short distances.  To be a bit more general, we call the energy scale above which the reduced symmetry occurs $M$, which can be, but does not necessarily need to be, the Planck scale, $M_{\rm pl}$.  For our effective approach to remain valid, $M$ should be above the inflationary Hubble scale, $H$.  Thus, $M$ lies within the range 
\begin{equation}
H < M \le M_{\rm pl} .
\label{HMMpl}
\end{equation}
As we shall only be evaluating the {\it leading\/} effects produced by the symmetry breaking, whenever possible we neglect corrections that are suppressed by the slow-roll parameters.  The only occasion we do not so is when $\dot\phi/H$ appears as an overall factor in the action for the fluctuations, which otherwise would vanish in this limit.  In this limit we have $a(t)=e^{Ht}$ and $H={\rm constant}$.

To see what would be the leading non-Gaussian signature produced by breaking the local Lorentz symmetry at short distances ($\ll 1/M$), we should include the least irrelevant operators (those of dimension five) that are cubic in the fluctuations, 
\begin{equation}
S_{\rm new} = \int dt\, a^3 {1\over M} {\dot\phi^3\over H^3} 
\int d^3\vec x\, \biggl\{ 
{1\over 6} d_1\, H^2\zeta^3 
- {1\over 2} {1\over a^2} d_2\, \zeta^2 \vec\nabla\cdot\vec\nabla\zeta 
\biggr\} . 
\label{newact}
\end{equation}
The $d_i$'s here are dimensionless constants.  Since we are working in the slowly rolling limit, we have neglected other, smaller terms as well as terms that are essentially the same, under an integration by parts, as those that we have already included.  Though $H$ is a constant in this limit, we have treated it as though it were an operator when counting the mass dimension of the first term, since it could have arisen from taking two derivatives of the metric, $R\propto H^2$.  Of course, a broken symmetry at short distances can be transmitted, through radiative corrections, to longer distances as well.  Here we simply assume that some principle restores the symmetry at larger distances---we are always free to add counterterms that precisely cancel any lower dimensional operators that are radiatively generated, so that the approach remains consistent, if a bit unnatural.

Evaluating the effect of these two operators on the three-point function, we find
\begin{eqnarray}
\langle 0(t)| \zeta(t,\vec x)\zeta(t,\vec y)\zeta(t,\vec z) |0(t)\rangle
&\!\!\!\!=\!\!\!\!& 
{1\over M} {\dot\phi^3\over H^3} 
\int {d^3\vec k_1\over (2\pi)^3} {d^3\vec k_2\over (2\pi)^3} 
{d^3\vec k_3\over (2\pi)^3}\, 
e^{i\vec k_1\cdot\vec x} e^{i\vec k_2\cdot\vec y} e^{i\vec k_3\cdot\vec z} 
(2\pi)^3\, \delta^3(\vec k_1 + \vec k_2 + \vec k_3)
\nonumber \\
&&
\times\int_{t_0}^t dt' 
\Bigl\{ d_1 H^2 e^{4Ht'} + d_2 e^{2Ht'}
(k_1^2 + k_2^2 + k_3^2) \Bigr\} 
\nonumber \\
&&\qquad\quad
i \bigl\{ 
\zeta_{k_1}(t) \zeta_{k_2}(t) \zeta_{k_3}(t) 
\zeta_{k_1}^*(t') \zeta_{k_2}^*(t') \zeta_{k_3}^*(t') 
\nonumber \\
&&\qquad\quad
- \zeta_{k_1}^*(t) \zeta_{k_2}^*(t) \zeta_{k_3}^*(t) 
\zeta_{k_1}(t') \zeta_{k_2}(t') \zeta_{k_3}(t') 
\bigr\} . 
\label{3pointuneval}
\end{eqnarray}
Thus far we left the limits on the time integral general; but the wavelengths relevant for cosmological observations are again those that have been stretched well outside the Hubble horizon before the end of the inflationary era, which corresponds to taking $t\to\infty$.\footnote{Actually, in the limit where we ignore all slowly rolling corrections to the Fourier modes $\zeta_k(t)$, there is a mild infrared divergence as we take $t\to\infty$.  This divergence disappears once we use a less idealized form for the modes and background.}

At the opposite end, at the initial limit of the time-integral which we have written as $t_{0}$, we can encounter a species of short-distance divergence directly related to the trans-Planckian---or in this instance, the ``trans-$M$''---regime.  In our effective approach, these apparent divergences are not necessarily genuine; instead, they signal the breakdown of the applicability of the effective description.  We make this connection with the trans-Planckian realm a little clearer by defining a cut-off momentum scale $k_\star$, which is a momentum exactly equal to the scale $M$ at the initial time $t_0$, 
\begin{equation}
k_\star = e^{Ht_0}M_{\rm pl} . 
\label{kstardef}
\end{equation}
The effective description remains consistent as long as we avoid momenta above this scale, so $k<k_\star$.

As a final step, we describe the effects of the symmetry-breaking operators by defining an effective, scale-dependent $f_{\rm nl}$ for each of the operators, 
\begin{eqnarray}
&&\!\!\!\!\!\!\!\!\!\!\!\!\!\!\!\!\!\!\!\!\!\!\!\!
\langle 0(t)| \tilde\zeta(t,\vec x) \tilde\zeta(t,\vec x) \tilde\zeta(t,\vec x) |0(t)\rangle\bigr|^{(i)} 
\nonumber \\
&&
\equiv - {3\over 5} {1\over\epsilon^2} {H^4\over M_{\rm pl}^4} \, 
\int {d^3\vec k_1\over (2\pi)^3} {d^3\vec k_2\over (2\pi)^3} 
     {d^3\vec k_3\over (2\pi)^3}\, 
e^{i\vec k_1\cdot\vec x} e^{i\vec k_2\cdot\vec y} e^{i\vec k_3\cdot\vec z} 
(2\pi)^3\, \delta^3(\vec k_1 + \vec k_2 + \vec k_3)
{k_1^3+k_2^3+k_3^3\over 8k_1^3k_2^3k_3^3} f_{\rm nl}^{(i)}(k_1,k_2,k_3) , 
\label{fnleff}
\end{eqnarray}
using the same form for the momentum integral as that which appeared when we first introduced $f_{\rm nl}$ for the toy fluctuations.  For the two operators that we have been studying, we find
\begin{eqnarray}
f_{\rm nl}^{(1)}(k_1,k_2,k_3) 
&\!\!\!\!=\!\!\!\!& 
- {5\over 3} {d_1\over 3\sqrt{2}} \sqrt{\epsilon} {M_{\rm pl}\over M} 
\biggl\{ {(k_1+k_2+k_3)(k_1^2+k_2^2+k_3^2)\over k_1^3+k_2^3+k_3^3} 
- {k_1k_2k_3\over k_1^3+k_2^3+k_3^3} \biggr\} + \cdots 
\nonumber \\ 
f_{\rm nl}^{(2)}(k_1,k_2,k_3) 
&\!\!\!\!=\!\!\!\!& 
- {5\over 3} 
{d_2\over\sqrt{2}} \sqrt{\epsilon} {M_{\rm pl}\over H} {1\over k_\star}
{k_1k_2k_3(k_1^2+k_2^2+k_3^2)\over (k_1+k_2+k_3)(k_1^3+k_2^3+k_3^3)} 
\sin\biggl( {M\over H} {k_1+k_2+k_3\over k_\star} \biggr) + \cdots ,
\label{efffnls}
\end{eqnarray}
to leading order in the slowly rolling parameters and neglecting terms that vanish in the $t_0\to -\infty$ limit, when possible.

Since the primordial non-Gaussianities have not been observed yet, it is perhaps still premature to present their detailed dependence on the momentum.  Therefore, we consider a couple characteristic cases.  When all of the momenta are nearly equal to each other, $k_i\sim k$, we have 
\begin{eqnarray}
f_{\rm nl}^{(1)}(k,k,k) 
&\!\!\!\!=\!\!\!\!& 
- {5\over 3} {8 d_1\over 9\sqrt{2}} \sqrt{\epsilon} {M_{\rm pl}\over M} + \cdots 
\nonumber \\ 
f_{\rm nl}^{(2)}(k,k,k) 
&\!\!\!\!=\!\!\!\!& 
- {5\over 3} {d_2\over 3\sqrt{2}} \sqrt{\epsilon} {M_{\rm pl}\over H}
{k\over k_\star}
\sin\biggl( 3 {M\over H} {k\over k_\star} \biggr) + \cdots . 
\label{efffnleq}
\end{eqnarray}
Another distinct case occurs when two are much larger than the third, $k_1\sim k_2\sim k \gg k_3$; we then have 
\begin{eqnarray}
f_{\rm nl}^{(1)}(k,k,k_3) 
&\!\!\!\!=\!\!\!\!& 
- {5\over 3} {2 d_1\over 3\sqrt{2}} \sqrt{\epsilon} {M_{\rm pl}\over M} 
+ \cdots 
\nonumber \\ 
f_{\rm nl}^{(2)}(k,k,k_3) 
&\!\!\!\!=\!\!\!\!& 
- {5\over 3} {d_2\over 2\sqrt{2}} \sqrt{\epsilon} {M_{\rm pl}\over H}
{k_3\over k_\star}
\sin\biggl( 2 {M\over H} {k\over k_\star} \biggr) + \cdots . 
\label{efffnlnl}
\end{eqnarray}
Note that the momenta are conserved, $\vec k_1 + \vec k_2 + \vec k_3 = 0$, which means that their vectors form a closed triangle.  The first case then corresponds to an equilateral triangle while the latter forms a narrow, isosceles one.  

Neglecting the numerical prefactors, and the model-dependent parameters $d_i$ which should be of the order of $1$, the size of the contribution of the first operator, $H^2\zeta^3$, to the three-point function is roughly 
\begin{equation}
\bigl| f_{\rm nl}^{(1)}\bigr| \sim
\sqrt{\epsilon} {M_{\rm pl}\over M} , 
\label{efffnl1}
\end{equation}
without changing very much with the shape of the triangle formed by the spatial momenta, $\vec k_i$.  The contribution from the second operator, $a^2\zeta^2\vec\nabla\cdot\vec\nabla\zeta$, however, is much smaller for a narrow isosceles triangle than for an equilateral one, where
\begin{equation}
\bigl| f_{\rm nl}^{(2){\rm equilateral}}\bigr| \sim
\sqrt{\epsilon} {M_{\rm pl}\over H} {k\over k_\star}
\sin\biggl( 3 {M\over H} {k\over k_\star} \biggr) . 
\label{efffnl2}
\end{equation}

Although none of the cosmological observations made so far are precise enough to be able to infer a non-vanishing value for the three-point function of the primordial fluctuations, their bounds already place quite strong limits on the scale $M$, indicating that it cannot be very far below the Planck scale.  For example, the most recent observations by the WMAP satellite suggest that for the two types of momentum triangles, the experimental bounds are \cite{wmap}
\begin{equation}
-151 < f_{\rm nl}^{\rm equilateral} < 253 
\qquad\qquad
-9 < f_{\rm nl}^{\rm isosceles} < 111 .
\label{wmapexp}
\end{equation}
So if the Lorentz symmetry were truly broken at very short distances, in a way that produces a fairly general set of symmetry-breaking operators, then for the first operator the scale associated with this breaking cannot be more than a couple orders of magnitude below the Planck scale $M_{\rm pl}$.  The second operator is even more constrained, since it is at least superficially proportional to $M_{\rm pl}/H$ and not $M_{\rm pl}/M$, although there is some dependence on the size of the $k$ associated with a particular measurement could have been at the ``initial'' time, $t_0$.

Comparing these results with the limits imposed on $M$ that can be inferred from the two-point function \cite{twopoint}, those from the three-point function are actually far more constraining.  The reason is a familiar one from particle physics experiments.  The corrections to the two-point function from the ``trans-Planckian'' regime (depending on what we assume about it) are typically suppressed, compared with the size of the standard inflationary prediction.  In contrast, the standard picture predicts a very small value for the non-Gaussian part of the primordial fluctuations, with the expected size for $f_{\rm nl}$ about the same as the slowly rolling parameters, whereas the generic ``trans-Planckian'' corrections can produce a potentially much larger $f_{\rm nl}$.  Sometimes, when hunting for the signals of new physics, it is more profitable to look not where the new signal would be the largest, but where the background is the smallest.

\begin{theacknowledgments}
The work described in this talk was done in collaboration with Rich Holman of Carnegie Mellon University \cite{nongauss}.  I am very grateful to the support provided by the Niels Bohr International Academy, and by the EU FP6 Marie Curie Research and Training Network ``UniverseNet'' (MRTN-CT-2006-035863) which provided funds for my travel to {\it Invisible Universe International Conference\/} held at the Palais de l'UNESCO in Paris.  
\end{theacknowledgments}

\bibliographystyle{aipproc}   

\begin{thebibliography}{99}

\bibitem{perts}
V.~F.~Mukhanov, H.~A.~Feldman and R.~H.~Brandenberger,
Phys.\ Rept.\  {\bf 215}, 203 (1992).

\bibitem{bardeen}
J.~M.~Bardeen,
Phys.\ Rev.\  D {\bf 22}, 1882 (1980).

\bibitem{transPlanck}
R.~H.~Brandenberger,
astro-ph/0411671; 
%
R.~H.~Brandenberger,
hep-ph/9910410.

\bibitem{picture}
R.~Brandenberger,
Phys.\ Today {\bf 61N3}, 44 (2008).

\bibitem{threepoint}
T.~J.~Allen, B.~Grinstein and M.~B.~Wise,
Phys.\ Lett.\  B {\bf 197}, 66 (1987); 
%
T.~Falk, R.~Rangarajan and M.~Srednicki,
Phys.\ Rev.\  D {\bf 46}, 4232 (1992); 
%
A.~Gangui, F.~Lucchin, S.~Matarrese and S.~Mollerach,
Astrophys.\ J.\  {\bf 430}, 447 (1994); 
%
V.~Acquaviva, N.~Bartolo, S.~Matarrese and A.~Riotto,
Nucl.\ Phys.\  B {\bf 667}, 119 (2003); 
%
J.~M.~Maldacena,
JHEP {\bf 0305}, 013 (2003). 

\bibitem{iscap}
K.~Schalm, G.~Shiu and J.~P.~van der Schaar,
JHEP {\bf 0404}, 076 (2004); 
%
B.~R.~Greene, K.~Schalm, G.~Shiu and J.~P.~van der Schaar,
JCAP {\bf 0502}, 001 (2005).

\bibitem{effstate}
H.~Collins and R.~Holman,
Phys.\ Rev.\  D {\bf 71}, 085009 (2005);
%
H.~Collins and R.~Holman,
hep-th/0507081. 

\bibitem{wmap}
E.~Komatsu {\it et al.}  [WMAP Collaboration],
Astrophys.\ J.\ Suppl.\  {\bf 180}, 330 (2009).

\bibitem{twopoint}
H.~Collins and R.~Holman,
Phys.\ Rev.\  D {\bf 77}, 105016 (2008). 

\bibitem{nongauss}
H.~Collins and R.~Holman,
Phys.\ Rev.\  D {\bf 80}, 043524 (2009).

\end{thebibliography}

\end{document}